\documentclass[letterpaper, 10 pt, conference]{ieeeconf}  

\IEEEoverridecommandlockouts                              

\usepackage{graphics} 
\usepackage{epsfig} 
\usepackage{mathptmx} 
\usepackage{times} 
\usepackage{amsmath} 
\usepackage{amssymb}  

\title{\LARGE \bf
Predicting Muscle Thickness Deformation from Muscle Activation Patterns: A Dual-Attention Framework
}

\author{Bangyu Lan$^{1}$ and Kenan Niu$^{1}$
\thanks{$^{1}$Bangyu Lan and Kenan Niu are with the Robotics and Mechatronics group, the Faculty of EEMCS at the University of Twente, P.O. Box 217, 7500 AE Enschede, Netherlands.
        {\tt\small bangyu@ieee.org (BL) and k.niu@utwente.nl (KN)}}%
}

\begin{document}

\maketitle
\thispagestyle{empty}
\pagestyle{empty}

\begin{abstract}
Understanding the relationship between muscle activation and thickness deformation is critical for diagnosing muscle-related diseases and monitoring muscle health. Although ultrasound technique can measure muscle thickness change during muscle movement, its application in portable devices is limited by wiring and data collection challenges. Surface electromyography (sEMG), on the other hand, records muscle bioelectrical signals as the muscle activation. This paper introduced a deep-learning approach to leverage sEMG signals for muscle thickness deformation prediction, eliminating the need for ultrasound measurement. Using a dual-attention framework combining self-attention and cross-attention mechanisms, this method predicted muscle deformation directly from sEMG data. Experimental results with six healthy subjects showed that the approach could accurately predict muscle excursion with an average precision of 0.923±0.900mm, which shows that this method can facilitate real-time portable muscle health monitoring, showing potential for applications in clinical diagnostics, sports science, and rehabilitation.

\end{abstract}

\section{INTRODUCTION}

Quantifying the relationship between muscle activation and thickness deformation is essential for understanding muscle dynamics and health \cite{engelke2018quantitative, campanini2020surface}, particularly in conditions such as Facioscapulohumeral Dystrophy \cite{an2018specific}. Traditional ultrasound imaging can visualize muscle thickness deformation. However, they are limited by their portability and the complexity to integrate into wearable devices \cite{molina2020validation, niu2024method, niu2018novel}. Surface electromyography (sEMG) device, which is portable and can capture muscle activation patterns, offers an alternative for describing the muscle health status. But it requires expert interpretation on the signals \cite{waris2018multiday, xue2019semg, chen2021non}. These challenges make it difficult to develop a practical, portable tool for simultaneously monitoring muscle thickness deformation and muscle activation. Additionally, other imaging solutions like MRI are impractical for real-time applications due to the cost and size \cite{corradini2019mr}.

\begin{figure}
    \centering
    \includegraphics[width=\linewidth]{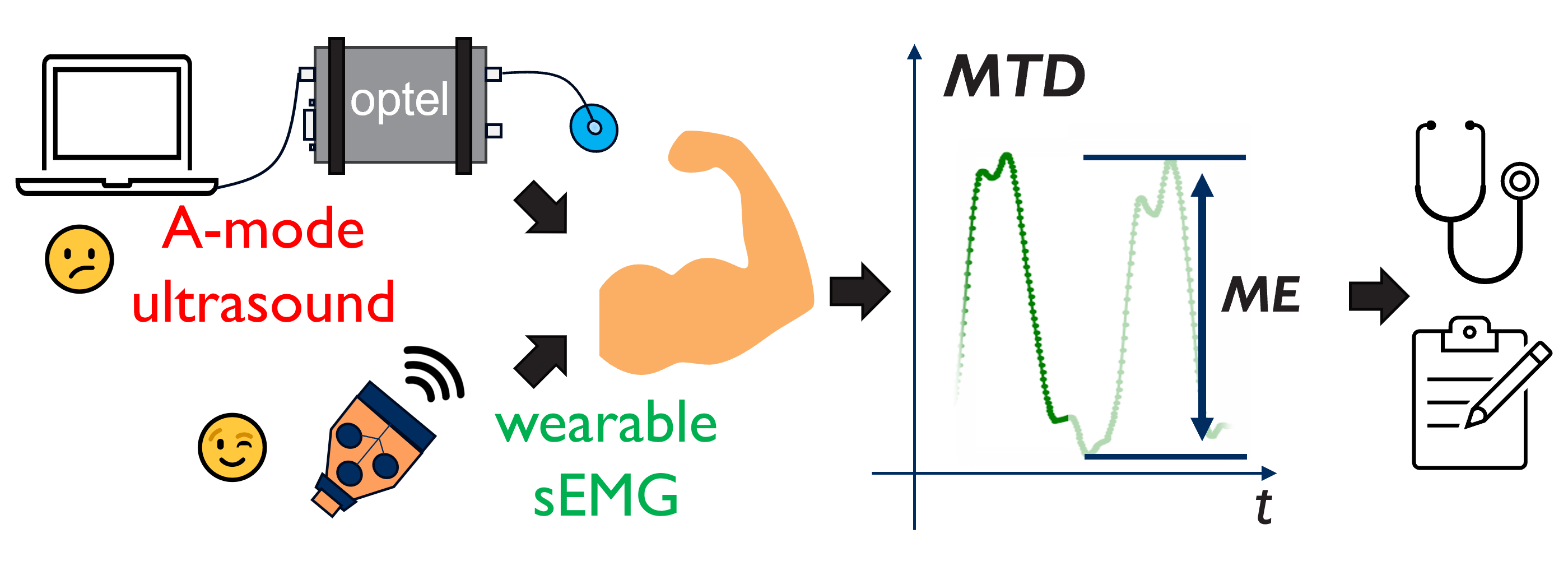}
    \caption{We proposed a method to use the sEMG to predict muscle
thickness contraction instead of the traditional ultrasound technique.
This approach could be used for disease diagnosis in daily life. (MTD:
muscle thickness deformation, ME: muscle excursion)}
    \label{fig:Graphical_Abstract}
\end{figure}

To address these limitations, we proposed an alternative solution that infers muscle thickness deformation quantitatively only from sEMG signals (see Fig. \ref{fig:Graphical_Abstract}). sEMG devices record muscle activities during contraction and extension \cite{waris2018multiday}, and the resulting signals exhibit specific patterns correlated with mechanical muscle behavior \cite{xue2019semg}. These observable patterns suggest a viable pathway for sEMG to infer muscle thickness changes \cite{cui2023high}, offering a portable and convenient tool for continuous tracking of muscle thickness deformation (MTD). From MTD, muscle excursion (ME) can be derived, providing a comprehensive view of the muscle health status, critical for diagnosing muscle diseases.

Our proposed method employs a novel dual-attention framework to correlate muscle activation with thickness deformation. This framework included hierarchical self-attention \cite{jia2020multi} and cross-attention \cite{pang2024cross} mechanisms. Self-attention captured long-range signal dependencies and dynamically adjusted the importance of different signal components \cite{vaswani2017attention}, while cross-attention merged and synthesized these features to provide comprehensive MTD information. This hierarchical combination enabled the model to accurately regress muscle contraction movements from the periodic features of the sEMG signals. Our experiment results with six subjects of different BMI indexes demonstrated the model's universality, generalizability, and robustness. The dual-attention framework's effectiveness was validated through ablation studies to assess the impact of the proposed loss function and the network architecture.

\section{METHOD AND MATERIAL}
The human-related muscle contraction experiment included six health subjects. Each subject performed three forearm motions repeatedly to generate a dataset to represent daily activities. They wore A-mode ultrasound and sEMG devices on their upper arm to record muscle thickness deformation and activation patterns. The ultrasound probe was positioned above the elbow, while the sEMG electrodes were attached to the middle upper arm. The subjects performed the motions in three stages, with the final stage involving a 500-gram weight. The collected data from all stages were used to train and evaluate the proposed network. The received ultrasound signals were processed to derive the muscle thickness deformation, while the synchronized sEMG data served as the input of the model.

The dual-attention network structure was designed to predict muscle thickness deformation from sEMG signals by capturing periodic patterns. The model used self-attention mechanisms to encode local muscle activation patterns from each sEMG channel, which were then analyzed by a cross-attention mechanism to predict muscle thickness deformation. The training process involved minimizing both the Mean Squared Error (MSE) loss and a muscle contraction loss. 

The method's generalizability was tested by training on data from four subjects and evaluating on the rest two "unseen" subjects (data has not been used in training). Domain adaptation was used later to finetune the model using only a small portion of data from the new subjects. Additional experiment was performed to validate model's robustness by involving a 500-gram weight. The evaluation metrics were related to the accuracy of muscle thickness deformation (MTD) and muscle excursion (ME).

\begin{table}[t!]
\caption{accuracy of evaluations for MTD and ME prediction}
\centering
\begin{tabular}{||c||c|c|c||}
\hline
Group & MTD(mm) & ME(mm) & ME(\%) \\
\hline
AD & 0.810 $\pm$ 0.826 & 0.743 $\pm$ 0.740 & 13.22 $\pm$ 14.59 \\
\hline
AB & 0.832 $\pm$ 0.859 & 0.806 $\pm$ 0.776 & 11.70 $\pm$ 10.82 \\
\hline
CD & 0.927 $\pm$ 0.888 & 0.745 $\pm$ 0.646 & 14.00 $\pm$ 14.30 \\
\hline
CB & 0.966 $\pm$ 0.983 & 0.824 $\pm$ 0.858 & 13.28 $\pm$ 15.94 \\
\hline
AF & 1.082 $\pm$ 1.309 & 0.961 $\pm$ 1.070 & 13.96 $\pm$ 16.93 \\
\hline
\end{tabular}
\label{tab:exp}
\end{table}

\section{RESULTS AND DISCUSSION}
The model demonstrated impressive performance in predicting MTD and ME, which was partly shown in TABLE \ref{tab:exp}. For MTD prediction, the average accuracy was 1.089$\pm$1.136 mm after domain adaptation, while for single-period ME prediction, the model achieved a more precise average accuracy of 0.923$\pm$0.900 mm and a relative accuracy of 14.03$\pm$15.15\% (w.r.t. ground truth). These results indicated that the model can achieve millimeter-level accuracy in tracking MTD and even sub-millimeter accuracy for ME regression, showcasing the effectiveness of the proposed approach.

The ablation study highlighted the benefits of the proposed components. The inclusion of the contraction loss function improved accuracy of around 4.2\%, demonstrating its effectiveness in enhancing model’s learning capability for ME regression. The cross-attention mechanism, while slightly decreased ME prediction accuracy, improved MTD tracking accuracy, demonstrating its importance to reduce phase and general shifts in the muscle deformation prediction. These enhancements showed that the dual-attention structure and the contraction loss function were critical for model's prediction.

The model’s robustness and adaptability were also validated. When the subjects carried a 500-gram weight, the model’s performance showed minimal degradation, maintaining accuracy within 1.3 to 1.4 mm. This indicated the robustness when the muscle strength changed. Furthermore, as the mentioned millimeter accuracy was achieved using only 20\% of the new subject’s data, the results demonstrated the model's quick adaptation capabilities and generalizability. These suggested that the model can efficiently generalize to different subjects and conditions, making it a promising tool for practical applications in muscle health monitoring.

\section{CONCLUSION}
This paper presented a dual-attention based structure to predict muscle thickness deformation (MTD) and muscle excursion (ME) from muscle activation signals (sEMG). The experiments demonstrated that the method was universal, generalizable, and robust across various subjects with different BMIs and ME ranges. The model's ability to quickly adapt to new individuals using limited data highlights its potential for broader applications. This work can significantly contribute to the development of portable and wearable medical devices, facilitating real-time and daily monitoring of patients' conditions.

\addtolength{\textheight}{-12cm}   


\bibliographystyle{IEEEtran}
\bibliography{IEEEfull}

\end{document}